\documentclass[a4paper,nofootinbib]{revtex4}
\usepackage{graphicx}
\usepackage{fancyhdr}
\usepackage{amsmath}
\usepackage{gensymb}
\def\be{\begin{equation}}
\def\ee{\end{equation}}
\def\ba{\begin{eqnarray}}
\def\ea{\end{eqnarray}}
\def\ra{\rightarrow}

\def\ltap{\;\centeron{\raise.35ex\hbox{$<$}}{\lower.65ex\hbox{$\sim$}}\;}
\def\gtap{\;\centeron{\raise.35ex\hbox{$>$}}{\lower.65ex\hbox{$\sim$}}\;}

\usepackage{color}
\usepackage{slashed}
\usepackage{epsfig, subfigure}
\usepackage{graphicx}
\usepackage{fancyhdr}
\usepackage{amsmath}
\begin{document}

\title{Large pseudo-scalar components in the C2HDM}

%

\author{Duarte Fontes}
    \email[E-mail: ]{duartefontes@tecnico.ulisboa.pt}
\affiliation{Departamento de F\'{\i}sica and CFTP, Instituto Superior T\'{e}cnico, Universidade de Lisboa,\\
Avenida Rovisco Pais 1, 1049-001 Lisboa, Portugal}
\author{Jorge C. Rom\~ao}
    \email[E-mail: ]{jorge.romao@tecnico.ulisboa.pt}
\affiliation{Departamento de F\'{\i}sica and CFTP, Instituto Superior T\'{e}cnico, Universidade de Lisboa,\\
Avenida Rovisco Pais 1, 1049-001 Lisboa, Portugal}
\author{Rui Santos}
    \email[E-mail: ]{rasantos@fc.ul.pt}
\affiliation{ISEL - 
 Instituto Superior de Engenharia de Lisboa,\\
 Instituto Polit\'ecnico de Lisboa 
 1959-007 Lisboa, Portugal}
\affiliation{Centro de F\'{\i}sica Te\'{o}rica e Computacional,
    Faculdade de Ci\^{e}ncias,
    Universidade de Lisboa,\\
    Campo Grande, Edif\'{\i}cio C8 1749-016 Lisboa, Portugal}

\author{Jo\~{a}o P.~Silva}
    \email[E-mail: ]{jpsilva@cftp.ist.utl.pt}
\affiliation{Departamento de F\'{\i}sica and CFTP, Instituto Superior T\'{e}cnico, Universidade de Lisboa,\\
Avenida Rovisco Pais 1, 1049-001 Lisboa, Portugal}

\begin{abstract}
We discuss the CP nature of the Yukawa couplings of the Higgs boson in the 
framework of a complex two Higgs doublet model (C2HDM). After analysing 
all data gathered during the Large Hadron Collider run 1,
the measurement of the Higgs couplings to the remaining SM particles already
restricts the parameter space of many extensions of the SM. However,
there is still room for very large CP-odd Yukawa couplings to light quarks
and leptons while the top-quark Yukawa coupling is already very constrained
by current data. Although indirect measurements of electric dipole moments 
play a very important role in constraining the pseudo-scalar components
of the Yukawa couplings, we argue that a direct measurement 
of the ratio of pseudoscalar to scalar couplings should be one of the
top priorities for the LHC run 2.
\end{abstract}

\maketitle

\thispagestyle{fancy}


\section{Introduction}

The Higgs boson discovery by the ATLAS~\cite{ATLASHiggs} and CMS~\cite{CMSHiggs} collaborations
at the Large Hadron Collider (LHC) has triggered a number of studies on multi-Higgs extension of the 
Standard Model (SM). Although the measured Higgs couplings show a very good agreement with
the SM predictions there is still room for interesting non-SM features to be explored at the
next LHC run. In fact, many multi-Higgs models provide interesting scenarios as is the
case of the complex two-Higgs double model (C2HDM). The 2HDM was proposed by T.~D.~Lee~\cite{Lee:1973iz}
as a means to explain the matter-antimatter asymmetry of the universe by allowing for an extra source
of CP-violation in the potential (see~\cite{hhg, ourreview} for a review).
The existing experimental data and in particular the one recently analysed at the LHC
has been used in several studies with the goal of constraining the parameter space of the
C2HDM~\cite{Barroso:2012wz, Inoue:2014nva,Cheung:2014oaa, Fontes:2014xva} or just
the Yukawa couplings~\cite{Brod:2013cka}. 

In this work we analyse C2HDM scenarios where the scalar component of the SM-like Higgs
Yukawa couplings to down-type quarks or to leptons vanish. The corresponding CP-odd
component has to be non-zero for the model to be in agreement with the LHC results. We will also
discuss situations where the Yukawa coupling is shared by the CP-even and CP-odd components
of the Higgs boson. 
Our approach is driven both by the current measurements and by the predictions
for the next LHC run. The
processes $pp \to h \to WW (ZZ)$, $pp \to h \to \gamma \gamma$ and
$pp \to h \to \tau^+ \tau^-$ are at present measured with an accuracy of about $20$\%.
The expected accuracies for the signal strengths of different Higgs
decay modes were presented by the ATLAS~\cite{ATLASpred} and CMS~\cite{CMSpred}
collaborations (see also~\cite{Dawson:2013bba}) for $\sqrt{s}=14$ TeV and for 300 and 3000 $fb^{-1}$
of integrated luminosities. The predictions for the signal strengths with the final states
$VV$, $\gamma \gamma$ and $\tau^+ \tau^-$ will be used to understand how the model will
perform at the end of the next LHC run because as shown in~\cite{Fontes:2014xva} they   
reproduce quantitatively the effect
of all possible final states in the Higgs decay. Therefore, the predicted accuracies
for the signal strength lead us to consider situations where, at $13$ TeV, the rates are
measured within either $10$\% or $5$\% of the SM prediction. We note that there is no visible  
difference in the plots when the energy is changed from $13$ to $14$ TeV 
as discussed in~\cite{Fontes:2014xva}.

\section{The C2HDM}

\label{sec:model}

The allowed parameter space of the C2HDM was recently reviewed in~\cite{Fontes:2015mea}
(see also~\cite{Fontes:2014xva, Ginzburg:2002wt, Khater:2003wq, ElKaffas:2007rq, ElKaffas:2006nt,
Grzadkowski:2009iz, Arhrib:2010ju, Barroso:2012wz}).
In this section we will briefly describe the C2HDM, a complex 2HDM 
with a softly broken $Z_2$ symmetry $\phi_1 \ra \phi_1, \phi_2 \ra -\phi_2$.
We write the scalar potential as~\cite{ourreview}
\ba
V_H
&=&
m_{11}^2 |\phi_1|^2
+ m_{22}^2 |\phi_2|^2
- m_{12}^2\, \phi_1^\dagger \phi_2
- (m_{12}^2)^\ast\, \phi_2^\dagger \phi_1
\nonumber\\*[2mm]
&&
+\, \frac{\lambda_1}{2} |\phi_1|^4
+ \frac{\lambda_2}{2} |\phi_2|^4
+ \lambda_3 |\phi_1|^2 |\phi_2|^2
+ \lambda_4\, (\phi_1^\dagger \phi_2)\, (\phi_2^\dagger \phi_1)
\nonumber\\*[2mm]
&&
+\, \frac{\lambda_5}{2} (\phi_1^\dagger \phi_2)^2
+ \frac{\lambda_5^\ast}{2} (\phi_2^\dagger \phi_1)^2 \, ,
\label{VH}
\ea
where all couplings except  $m_{12}^2$ and $\lambda_5$ are real due to the hermiticity
of the potential and $\textrm{arg}(\lambda_5) \neq 2\, \textrm{arg}(m_{12}^2)$, 
so that the two phases cannot be removed simultaneously~\cite{Ginzburg:2002wt}.

We work in a basis where the vacuum expectation values (vevs) are real. The corresponding
CP-conserving 2HDM, is obtained from the C2HDM by taking $m_{12}^2$ and $\lambda_5$ real.
Defining the scalar doublets as
\be
\phi_1 =
\left(
\begin{array}{c}
\varphi_1^+\\
\tfrac{1}{\sqrt{2}} (v_1 + \eta_1 + i \chi_1)
\end{array}
\right),
\hspace{5ex}
\phi_2 =
\left(
\begin{array}{c}
\varphi_2^+\\
\tfrac{1}{\sqrt{2}} (v_2 + \eta_2 + i \chi_2)
\end{array}
\right),
\ee
with $v = \sqrt{v_1^2 + v_2^2} = (\sqrt{2} G_\mu)^{-1/2} = 246$ GeV, they can be 
written in the Higgs basis as~\cite{LS,BS}
\be
\left(
\begin{array}{c}
H_1\\
H_2
\end{array}
\right)
=
\left(
\begin{array}{cc}
c_{\beta} & s_{\beta}\\
- s_{\beta} & c_{\beta}
\end{array}
\right)
\left(
\begin{array}{c}
\phi_1\\
\phi_2
\end{array}
\right),
\ee
where $\tan{\beta} = v_2/v_1$,
$c_\beta = \cos{\beta}$, and $s_\beta = \sin{\beta}$.
In the Higgs basis the second doublet does not get a vev
and the Goldstone bosons are in the first doublet.

Defining $\eta_3$ as the neutral imaginary component of the $H_2$ doublet, 
the mass eigenstates are obtained from
the three neutral states via the rotation matrix $R$
\be
\left(
\begin{array}{c}
h_1\\
h_2\\
h_3
\end{array}
\right)
= R
\left(
\begin{array}{c}
\eta_1\\
\eta_2\\
\eta_3
\end{array}
\right)
\label{h_as_eta}
\ee
which will diagonalize the mass matrix of the neutral states
via
\be
R\, {\cal M}^2\, R^T = \textrm{diag} \left(m_1^2, m_2^2, m_3^2 \right),
\ee
and $m_1 \leq m_2 \leq m_3$ are the masses of the neutral Higgs particles.
We parametrize the mixing matrix $R$ as \cite{ElKaffas:2007rq}
\be
R =
\left(
\begin{array}{ccc}
c_1 c_2 & s_1 c_2 & s_2\\
-(c_1 s_2 s_3 + s_1 c_3) & c_1 c_3 - s_1 s_2 s_3  & c_2 s_3\\
- c_1 s_2 c_3 + s_1 s_3 & -(c_1 s_3 + s_1 s_2 c_3) & c_2 c_3
\end{array}
\right)
\label{matrixR}
\ee
with $s_i = \sin{\alpha_i}$ and
$c_i = \cos{\alpha_i}$ ($i = 1, 2, 3$) and
\be
- \pi/2 < \alpha_1 \leq \pi/2,
\hspace{5ex}
- \pi/2 < \alpha_2 \leq \pi/2,
\hspace{5ex}
- \pi/2 \leq \alpha_3 \leq \pi/2.
\label{range_alpha}
\ee

We choose the 9 independent parameters of the C2HDM to be $v$, $\tan \beta$, $m_{H^\pm}$,
$\alpha_1$, $\alpha_2$, $\alpha_3$, $m_1$, $m_2$, and $\textrm{Re}(m_{12}^2)$.
The mass of the heavier neutral scalar is a dependent parameter given by
\be
m_3^2 = \frac{m_1^2\, R_{13} (R_{12} \tan{\beta} - R_{11})
+ m_2^2\ R_{23} (R_{22} \tan{\beta} - R_{21})}{R_{33} (R_{31} - R_{32} \tan{\beta})}.
\label{m3_derived}
\ee
The parameter space will be constrained by the condition $m_3 > m_2$.

The Higgs coupling to gauge bosons is~\cite{Barroso:2012wz}
\be
C = c_\beta R_{11} + s_\beta R_{12}
=
\cos{(\alpha_2)}\, \cos{(\alpha_1 - \beta)} \, .
\label{C}
\ee
Regarding the Yukawa couplings, the $Z_2$ symmetry is extended 
to the Yukawa Lagrangian~\cite{GWP} to avoid flavour changing neutral currents (FCNC).
The up-type quarks couple to $\phi_2$ and the usual four models
are obtained by coupling down-type quarks and charged leptons
to $\phi_2$ (Type I) or to $\phi_1$ (Type II); or by coupling
the down-type quarks to $\phi_1$ and the charged leptons
to $\phi_2$ (Flipped) or finally by coupling the down-type
quarks to $\phi_2$ and the charged leptons to $\phi_1$ (Lepton Specific).
The Yukawa couplings can then be written, relative to the SM ones, as $a + i b \gamma_5$ with
the coefficients presented in table~\ref{tab:1}.
\begin{table}[h!]
\centering
\begin{tabular}{lcccccccc}
\hline
 & & Type I  & & Type II & & Lepton & & Flipped \\
 & & & & & & Specific & & \\
\hline 
Up  & &
$\tfrac{R_{12}}{s_{\beta}} - i c_\beta \tfrac{R_{13}}{s_{\beta}} \gamma_5$   & &
$\tfrac{R_{12}}{s_{\beta}} - i c_\beta \tfrac{R_{13}}{s_{\beta}} \gamma_5$  & &
$\tfrac{R_{12}}{s_{\beta}} - i c_\beta \tfrac{R_{13}}{s_{\beta}} \gamma_5$   & &
$\tfrac{R_{12}}{s_{\beta}} - i c_\beta \tfrac{R_{13}}{s_{\beta}} \gamma_5$  \\*[2mm]
\hline
Down  & &
$\tfrac{R_{12}}{s_{\beta}} + i c_\beta \tfrac{R_{13}}{s_{\beta}} \gamma_5$   & &
$\tfrac{R_{11}}{c_{\beta}} - i s_\beta \tfrac{R_{13}}{c_{\beta}} \gamma_5$    & &
$\tfrac{R_{12}}{s_{\beta}} + i c_\beta \tfrac{R_{13}}{s_{\beta}} \gamma_5$   & &
$\tfrac{R_{11}}{c_{\beta}} - i s_\beta \tfrac{R_{13}}{c_{\beta}} \gamma_5$    \\*[2mm]
\hline
Leptons  & &
$\tfrac{R_{12}}{s_{\beta}} + i c_\beta \tfrac{R_{13}}{s_{\beta}} \gamma_5$   & &
$\tfrac{R_{11}}{c_{\beta}} - i s_\beta \tfrac{R_{13}}{c_{\beta}} \gamma_5$    & &
$\tfrac{R_{11}}{c_{\beta}} - i s_\beta \tfrac{R_{13}}{c_{\beta}} \gamma_5$   & &
$\tfrac{R_{12}}{s_{\beta}} + i c_\beta \tfrac{R_{13}}{s_{\beta}} \gamma_5$   \\*[2mm]
\hline
\end{tabular}
\caption{\label{tab:1} Yukawa couplings of the lightest scalar, $h_1$,
in the form $a + i b\gamma_5$.}
\end{table}

\section{Results and Discussion}

In order to perform our analysis we generate points in parameter space 
in the following intervals: the lightest neutral scalar is $m_1 = 125$ GeV~\footnote{The
latest results on the measurement of the Higgs mass are $125.36 \pm 0.37$ GeV
from ATLAS~\cite{Aad:2014aba} and $125.02 +0.26 -0.27$ (stat) $+0.14 -0.15$ (syst) GeV
from CMS~\cite{Khachatryan:2014jba}.},
the angles $\alpha_{1,2,3}$ all vary in the interval $[-\pi/2, \, \pi/2]$,
$1 \leq \tan{\beta} \leq 30$, $ m_1 \leq m_2 \leq 900\, \textrm{GeV}$ and
$-(900\, \textrm{GeV})^2 \leq Re(m_{12}^2) \leq (900\, \textrm{GeV})^2$.
The points are generated randomly subject to the following constraints,
\begin{itemize}
\item
\textbf{B-physics} - $b \ra s \gamma$, in Type II/F 
we choose the range for the charged Higgs mass as
$340\, \textrm{GeV} \leq m_{H^\pm} \leq 900\, \textrm{GeV}$~\cite{BB},
while in Type I/LS the range is $100\, \textrm{GeV} \leq m_{H^\pm} \leq 900\, \textrm{GeV}$.
The remaining constraints from
B-physics~\cite{Deschamps:2009rh, gfitter1} (and from the
$R_b\equiv\Gamma(Z\to b\bar{b})/\Gamma(Z\to{\rm hadrons})$~\cite{Ztobb} measurement)
force $\tan{\beta} \gtrsim 1$ for all models;
\item
\textbf{LEP} - The charged Higgs mass is above 100 GeV 
due to LEP searches on $e^+ e^- \to H^+ H^-$~\cite{Abbiendi:2013hk}
(we also consider the LHC results on $pp \to \bar t \, t (\to H^+ \bar b )$~\cite{ATLASICHEP, CMSICHEP}).
Very light neutral scalars are also constrained by LEP results~\cite{lepewwg};
\item
\textbf{LHC - bounds on heavy scalars} - The most relevant searches
for the C2HDM are $pp \to \phi \to W^+ W^- (ZZ)$~\cite{ATLAS:2012ac, Chatrchyan:2013yoa}
and $pp \to \phi \to \tau^+ \tau^-$~\cite{Aad:2014vgg,
Khachatryan:2014jya}, where $\phi$ is a spin zero particle;
\item
\textbf{Theoretical constraints} - the potential is bounded from
below~\cite{Deshpande:1977rw}, perturbative unitarity is
enforced~\cite{Kanemura:1993hm, Akeroyd:2000wc,Ginzburg:2003fe} and all allowed points conform
to the oblique radiative parameters~\cite{Peskin:1991sw, Grimus:2008nb, Baak:2012kk}.
\end{itemize}

Finally we consider the results stemming from the the 125 GeV Higgs couplings measurements.
The signal strength is defined as
\begin{equation}
\mu^{h_i}_f \, = \, \frac{\sigma \, {\rm BR} (h_i \to
  f)}{\sigma^{\scriptscriptstyle {\rm SM}} \, {\rm BR^{\scriptscriptstyle{\rm SM}}} (h_i \to f)}
\label{eg-rg}
\end{equation}
where $\sigma$ is the Higgs boson production cross section and ${\rm BR} (h_i \to f)$ is
the branching ratio of the $h_i$ decay into the final state $f$;  $\sigma^{\scriptscriptstyle {\rm {SM}}}$
and ${\rm BR^{\scriptscriptstyle {\rm SM}}}(h \to f)$ are the respective quantities calculated in the SM.
Values for the cross sections were obtained from: HIGLU~\cite{Spira:1995mt} - gluon fusion at NNLO, 
together with the corresponding expressions for the CP-violating model in~\cite{Fontes:2014xva};
SusHi~\cite{Harlander:2012pb} - $b \bar{b} \ra h$ at NNLO; \cite{LHCCrossSections} - $Vh$ (associated production), 
$t \bar{t} h$ and $VV \ra h$ (vector boson fusion). 
As previously discussed, we will force $\mu_{VV}$, $\mu_{\gamma \gamma}$ and $\mu_{\tau \tau}$ to be within $20$\%
of the expected SM value, which at present roughly matches the average precision at $1\sigma$. 
Taking all other processes into account has no significant impact on the results as shown 
in~\cite{Fontes:2014xva}.

\begin{figure}[h!]
\centering
\includegraphics[width=6.3in,angle=0]{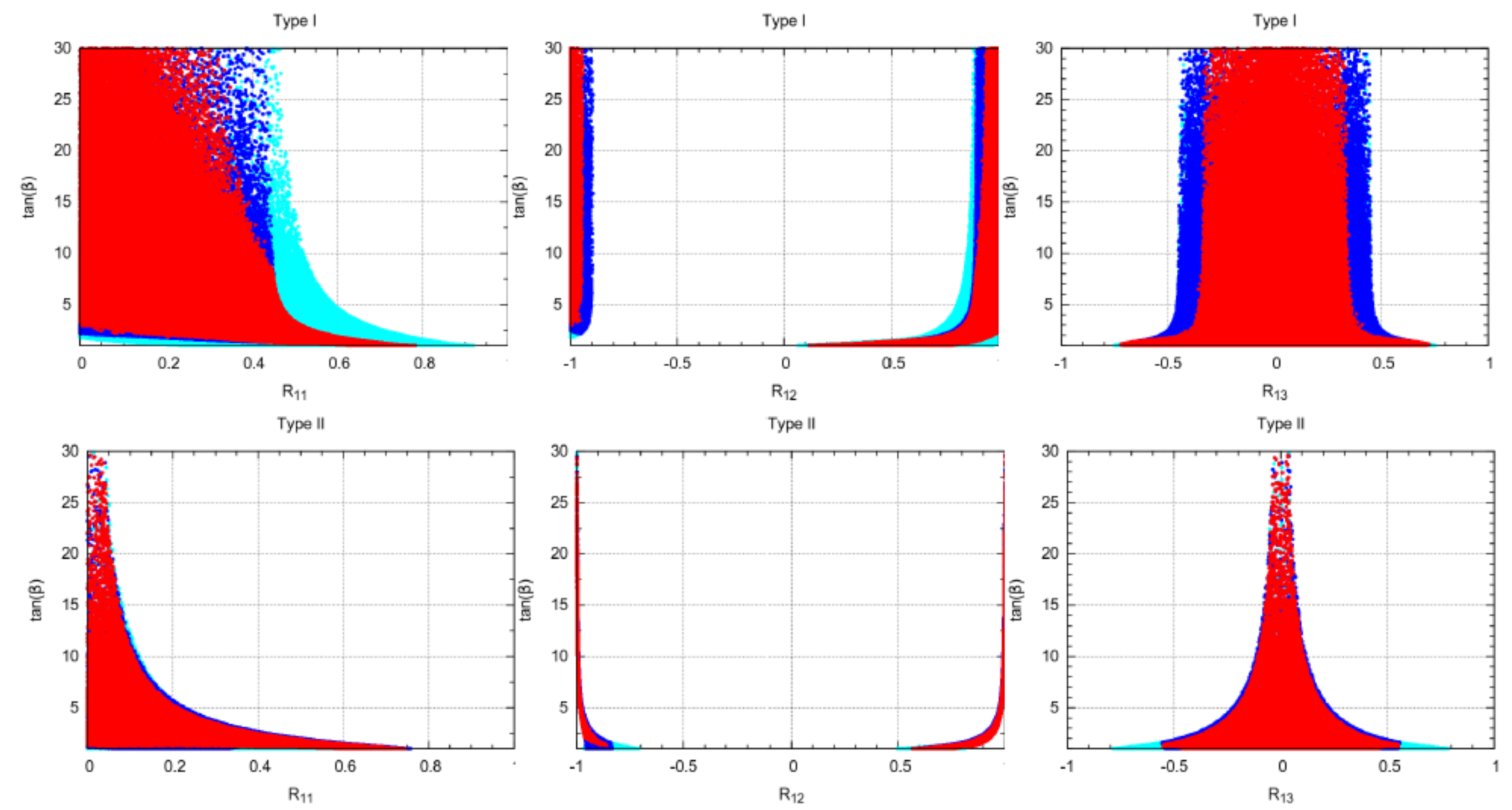}
\vspace{-0.3cm}
\caption{Top: $\tan \beta$ as a function of $R_{11}$ (left), $R_{12}$ (middle)
and $R_{13}$ for Type I. Bottom: same but for Type II. The rates are taken to be
within 20$\%$ of the SM predictions. The colours are superimposed with
cyan/light-grey for $\mu_{VV}$, blue/black for $\mu_{\tau \tau}$ and finally red/dark-grey for $\mu_{\gamma \gamma}$ with
$\sqrt{s} = 8$ TeV.}
\label{fig-1}
\end{figure}

In the C2HDM there is only one way to obtain pure scalar states. 
When we set $s_2=0$ we get $R_{13}=0$ and all pseudoscalar components
vanish. However, depending on the model type, there are in principle two ways to
obtain a vanishing scalar component. One is by setting $R_{12} = 0$.
However, as shown in figure~\ref{fig-1} (middle), values of $R_{12} \approx 0$ are
excluded when all constraints are taken into account.
The other possibility is to have $R_{11}= c_1 \, c_2 = 0$ which is still
allowed as shown in figure~\ref{fig-1} (left). This can be obtained by setting
either $c_2=0$ or $c_1=0$. $c_2=0$ is excluded as it would mean $g_{h_1 VV}=0$, 
where $V$ is a massive vector boson. Finally we can choose $c_1=0$. The values 
for $a_{F}$ and $b_F$ ($F=U,D,L$) in this scenario are presented in Table~\ref{tab:2}.

\begin{table}[h!]
\centering
\begin{tabular}{lcccccccc}
\hline
\hline
Type I  & &
$a_U=a_D=a_L=\tfrac{c_2}{s_{\beta}}$  & &
$b_U=- b_D=-b_L=-\tfrac{s_2}{t_{\beta}}$   \\*[2mm]
\hline 
Type II  & &
$a_D = a_L=0$  & &
$b_D = b_L = - s_2 \, t_{\beta}$  \\*[2mm]
\hline
Type F  & &
$a_D = 0$   & &
$b_D = - s_2 \, t_{\beta}$    \\*[2mm]
\hline
Type LS  & &
$a_L=0$   & &
$b_L = - s_2 \, t_{\beta}$   \\*[2mm]
\hline
\end{tabular}
\caption{\label{tab:2} $a_F$ and $b_F$ limits for $c_1 = 0$ for the four model types.}
\end{table}

As shown in table~\ref{tab:2} the scenarios where the scalar component vanishes
arise only in models Type II, F and LS. In Type II one can have $a_D = a_L = 0$ 
while in F (LS) only $a_D = 0$ ($a_L = 0$) is possible. In this scenario the 
coupling to gauge bosons is 
\begin{equation}
C^2=s_\beta^2 c_2^2 \, .
\end{equation}
Note that even if $s_2=0$ the pseudoscalar component can still be large due to
a large value of $\tan \beta$.

\begin{figure}[h!]
\centering
\includegraphics[width=6.3in,angle=0]{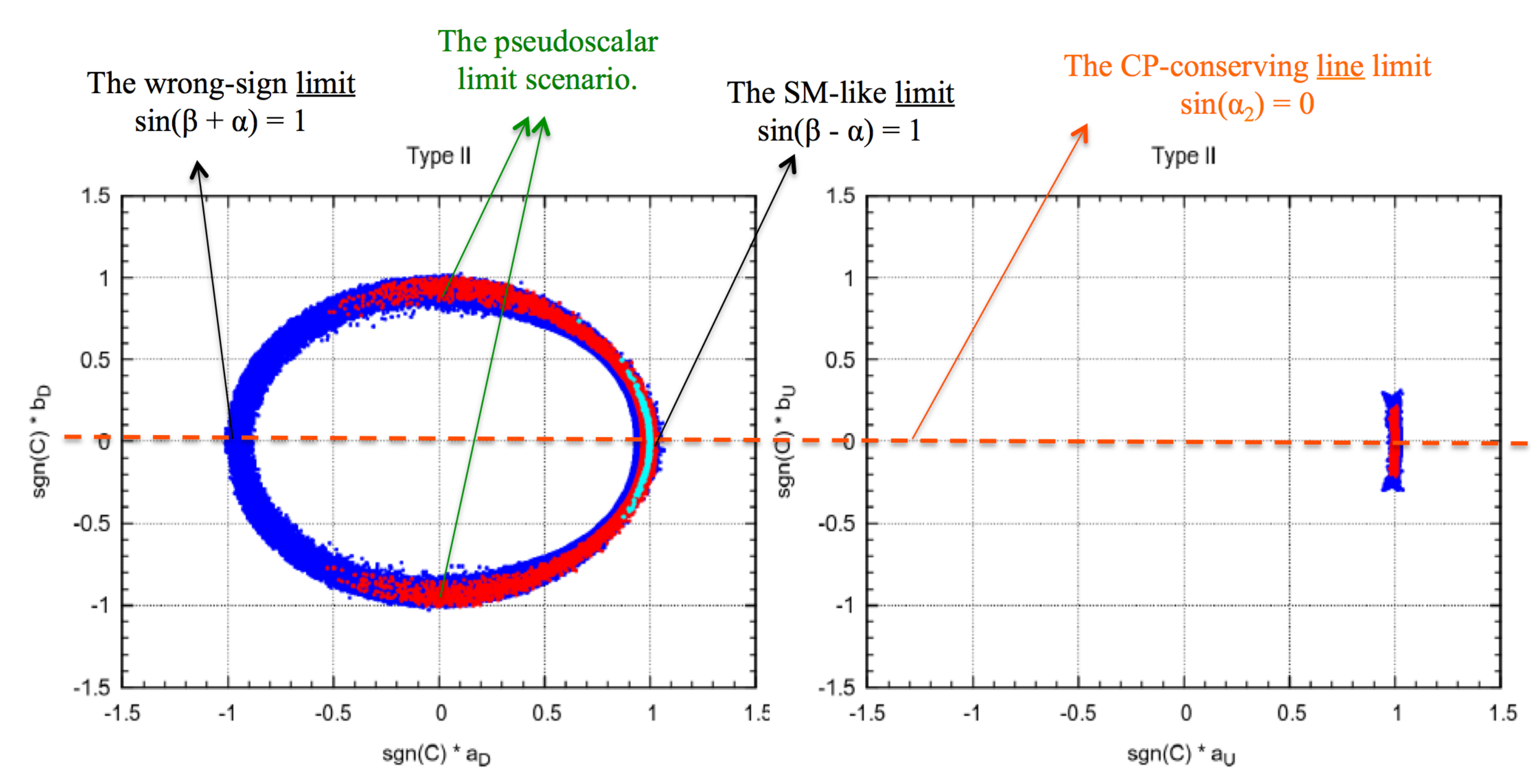}
\vspace{-0.3cm}
\caption{Left: sgn$(C)$ $b_D$ $=$ sgn$(C)$ $b_L$ as a function of sgn$(C)$ $a_D$ $=$ sgn$(C)$ $a_L$  for Type II and a center of mass
energy of $13$ TeV with all rates
at $10$\% (blue/black), $5$\% (red/dark-grey), and $1$\% (cyan/light-grey).
Right: same, but for sgn$(C)$ $b_U$ as a function of sgn$(C)$ $a_U$.}
\label{fig0}
\end{figure}

We will now discuss in detail the allowed parameter space in the $(a_F, b_F)$  plane 
for the different model types. We will plot sgn$(C) \,a_F$ (sgn$(C) \,b_F$) 
instead of $a_F$ ($b_F$) with $F=U,D,L$, to avoid the dependence
on the phase conventions in choosing the range for the angles $\alpha_i$.
In the left panel of figure~\ref{fig0} we show $b_D=b_L$ as a function
of $a_D=a_L$ for Type II and $\sqrt{s} = 13$ TeV with all rates
at $10$\% (blue/black), $5$\% (red/dark-grey), and $1$\% (cyan/light-grey).
We start by noting that this scenario is still possible with the rates within 
$5$\% of the SM value at the LHC and  at $\sqrt{s} = 13$ TeV. This scenario
can only be excluded by a measurement of the rates if the accuracy reaches  
about $1$\%. The constraints on the model force $|b_D| \to 1$ when $|a_D| \to 0$.
When $|b_D| \approx 1$, the couplings of the up-type quarks to the lightest
Higgs have the form
\begin{equation}
a_U^2=( 1 -s_2^4)=(1-1/t_\beta^4), \quad b_U^2=s_2^4=1/t^4_\beta,
\end{equation}
while the coupling to massive gauge bosons is now
\begin{equation}
C^2 = (t^2_\beta-1)/(t^2_\beta+1) = (1 - s_2^2)/(1 + s_2^2)  \, .
\end{equation}
In the right panel of figure~\ref{fig0} we show $b_U$ as a function
of $a_U$ for Type II with the same colour code.
We conclude from the plot that the constraint on the values of $(a_U,\, b_U)$ are already quite
strong and will be much stronger in the future just taking into account the
measurement of the rates.

\begin{figure}[h!]
\centering
\includegraphics[width=6.3in,angle=0]{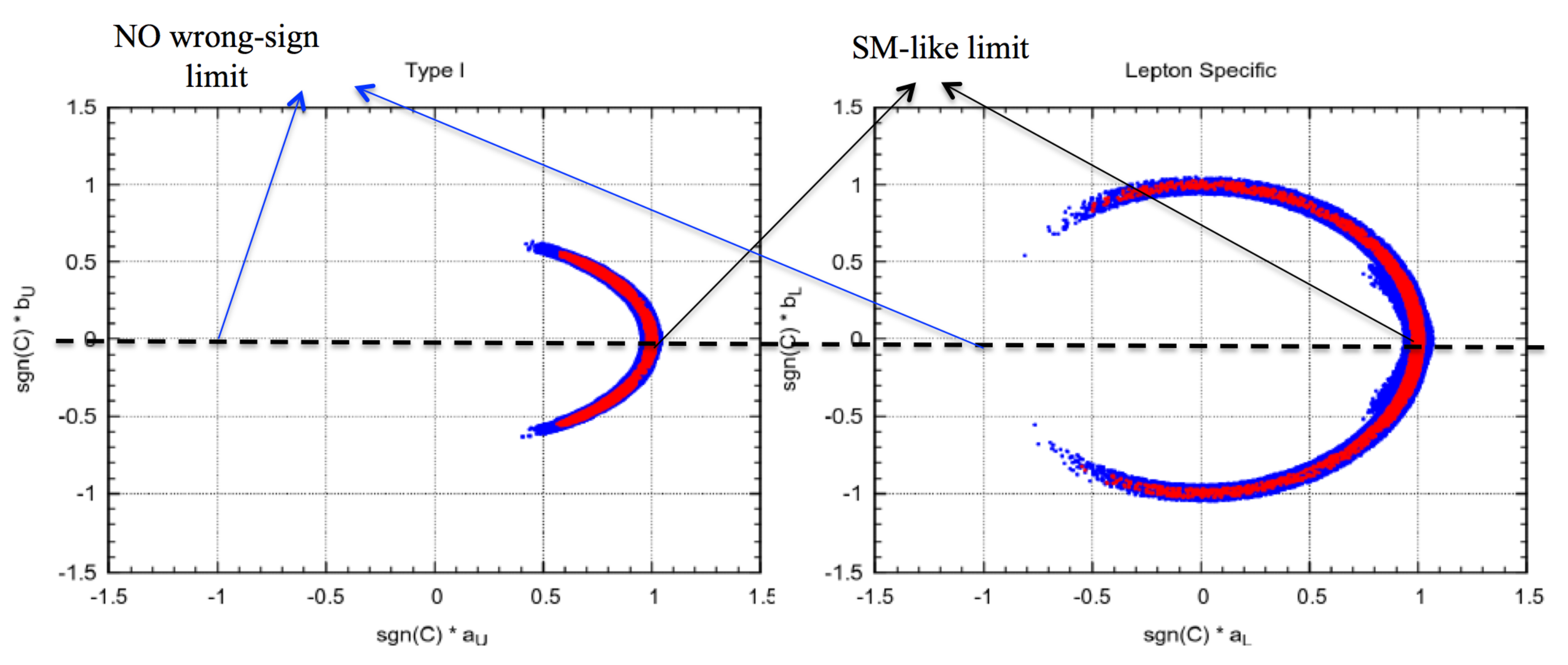}
\vspace{-0.3cm}
\caption{Left: sgn$(C)$ $b_U$ as a function of sgn$(C)$ $a_U$ for Type I and a center of mass
energy of $13$ TeV with all rates
at $10$\% (blue/black) and $5$\% (red/dark-grey).
Right: sgn$(C)$ $b_L$ as a function of sgn$(C)$ $a_L$ for LS and a center of mass
energy of $13$ TeV with all rates
at $10$\% (blue/black) and $5$\% (red/dark-grey).}
\label{fig1}
\end{figure}

In the left panel of figure~\ref{fig1} we show $b_U$ as a function of
$a_U$ for Type I and $\sqrt{s}=13$ TeV with all rates
at $10$\% (blue/black) and $5$\% (red/dark-grey).
We should point out that even at $10$\%
there are still allowed points close to $(a,b)=(0.5,0.6)$
with no dramatic changes occurring for an increase in accuracy to $5$\%. In the right
panel we present $b_L$ as a function of $a_L$ for Type LS with the same colour code.
Here again the $(a_L,b_L)=(0,1)$ scenario is still
allowed with both $10$\% and $5$\% accuracy. However, as was previously shown,
the wrong sign limit is not allowed for the LS model~\cite{Fontes:2014tga, Ferreira:2014dya}.
Nevertheless, in the C2HDM, the scalar component sgn$(C) \, a_L$
can  reach values close to $-0.8$. Finally, for the up-type and down-type quarks, the plots are very similar
to the one in the right panel of figure~\ref{fig0} for Type II.

\begin{figure}[h!]
\centering
\includegraphics[width=6.3in,angle=0]{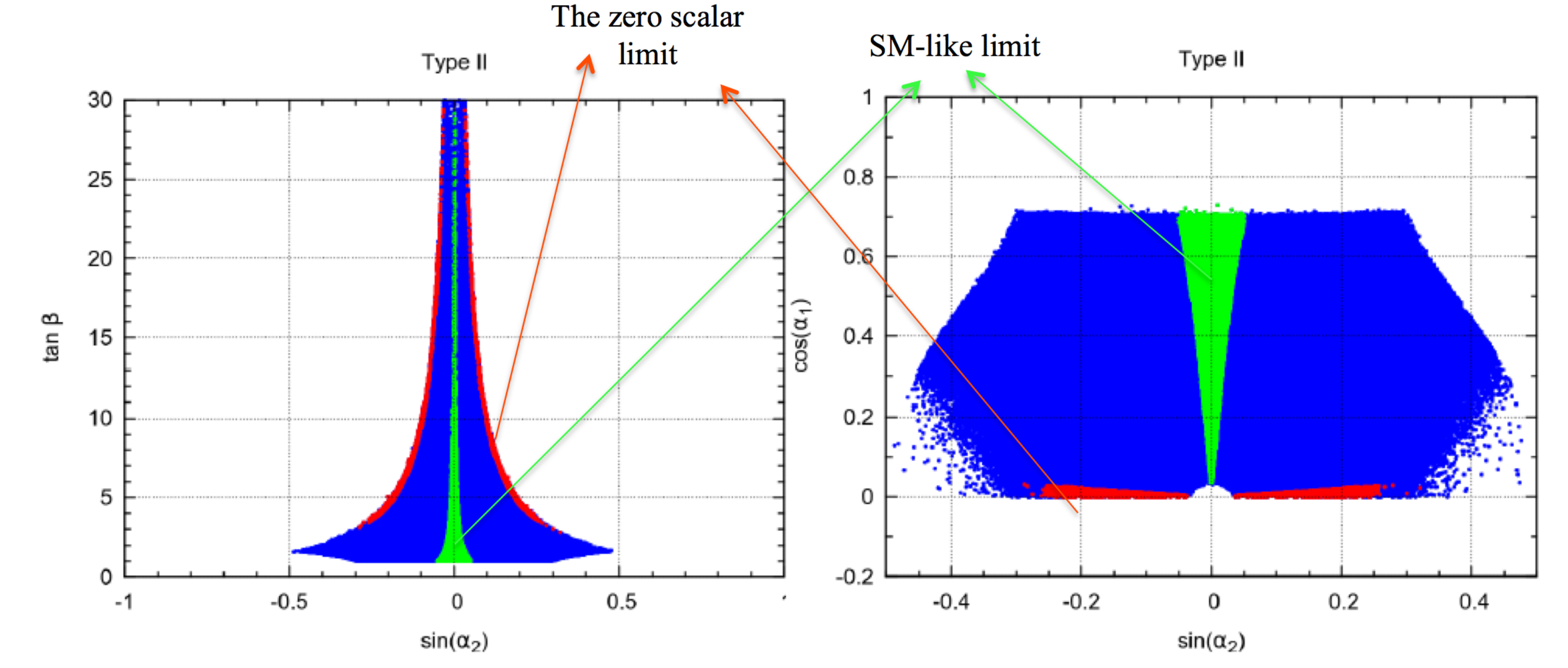}
\vspace{-0.3cm}
\caption{Left: $\tan \beta$ as a function of $\sin \alpha_2$ for Type II and a center of mass
energy of $13$ TeV with all rates
at $10$\% (blue/black). In red/dark-grey we show the points with $|a_D| < 0.1$ and $||b_D|-1| < 0.1$ and
in green $|b_D| < 0.05$ and $||a_D|-1| < 0.05$
Right: same, with $\tan \beta$ replaced by $\cos \alpha_1$.}
\label{fig2}
\end{figure}

In the left panel of figure~\ref{fig2} we present the allowed space in the $\sin \alpha_2$-$\tan{\beta}$ plane,
for Type II and $\sqrt{s} = 13$ TeV. Rates at $10$\% are shown in blue/black while in red/dark-grey we present
the points with $|a_D| < 0.1$ and $||b_D| - 1| < 0.1$ and
in green $|b_D| < 0.05$ and $||a_D| - 1| < 0.05$. The right panel now shows the allowed
space in the $\sin \alpha_2$-$\cos \alpha_1$ plane. The purpose of these plots is to
pinpoint the main differences between the SM-like scenario, where $(|a_D|,|b_D|) \approx (1,0)$ and 
the pseudoscalar scenario where $(|a_D|,|b_D|) \approx (0,1)$.
In the SM-like scenario $\sin \alpha_2 \approx 0$,
$\tan \beta$ is not constrained and the allowed values of $\sin \alpha_2$ grow with increasing
$\cos \alpha_1$. In the pseudoscalar scenario $\cos \alpha_1 \approx 0$, $\sin \alpha_2$
and $\tan \beta$ are strongly correlated and $\tan \beta$ has to be above $\approx 3$.
We note that all values of $a_D$ and $b_D$ are allowed provided $a_D^2 + b_D^2 \approx 1$.

\subsection{Direct measurements of the CP-violating angle}

We have seen that the precise measurements of the Higgs couplings
allows us to constrain both the scalar and the pseudoscalar 
Yukawas in the C2HDM. However, a direct measurement of the relative size of the pseudoscalar
to scalar coupling is important because it directly probes the Higgs couplings
to light quarks and leptons. Moreover, when combined with EDMs it can provide universality
tests for the CP-odd components of the Yukawas.  

The angle that measures the pseudoscalar to scalar ratio, $\phi_i$, is defined by
\begin{equation}
\tan \phi_i  = b_i/a_i \qquad i=U,\, D, \, L \, ,
\end{equation}
and could in principle be measured for all Yukawa couplings. 
Direct measurements of this ratio in the up-quark sector, $b_U/a_U$,
was first proposed in~\cite{Gunion:1996xu} and more recently in~\cite{Ellis:2013yxa, He:2014xla, Boudjema:2015nda}.
The process $pp \to hjj$~\cite{DelDuca:2001fn} also allows to probe the same vertex
as discussed in~\cite{Field:2002gt, Dolan:2014upa}.
In reference~\cite{Dolan:2014upa} an exclusion of $\phi_t > 40 \degree$ ($\phi_t > 25 \degree$)
for a luminosity of 50 fb$^{-1}$ (300 fb$^{-1}$) was obtained for 14 TeV
and assuming $\phi_t=0$ as the null hypothesis. A study of the $\tau^+ \tau^- h$
vertex was proposed in~\cite{Berge:2008wi} (see also~\cite{Harnik:2013aja, Askew:2015mda}) 
and a detailed study taking into account the
main backgrounds~\cite{Berge:2014sra} lead to an estimate in the precision
of $\Delta \phi_\tau$ of $27 \degree$ ($14.3 \degree$) for a luminosity of 150 fb$^{-1}$
(500 fb$^{-1}$) and $\sqrt{s}=14$ Tev. 

The number of independent measurements of $\phi_i$ one needs depends on the C2HDM Yukawa type.
For Type I one process is enough since $\phi_U = \phi_D = \phi_L$. For all other types
we need two independent measurements. For type II and LS the planned measurements of $\phi_t$ and
$\phi_\tau$ would be enough while for type F we would need $\phi_b$. Incompatibility in the
measured values of $\phi_t$ and $\phi_\tau$ would exclude both Type I and F.  

\begin{figure}[h!]
\centering
\includegraphics[width=6.3in,angle=0]{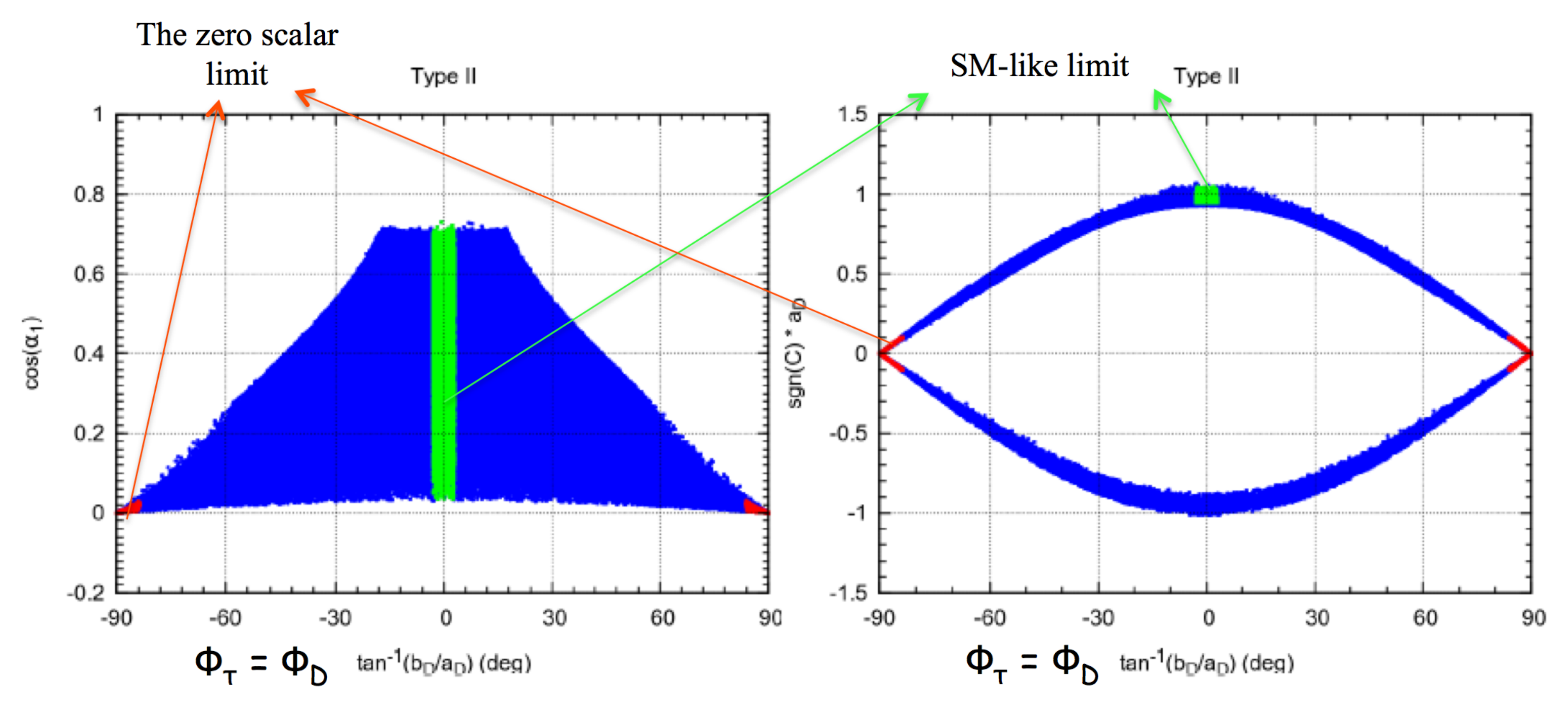}
\vspace{-0.3cm}
\caption{Left: $\cos \alpha_1$ as a function of $\tan^{-1} (b_D/a_D)$ for Type II and a center of mass
energy of $13$ TeV with all rates
at $10$\% (blue/black). In red/dark-grey we show the points with $|a_D| < 0.1$ and $||b_D|-1| < 0.1$ and
in green $|b_D| < 0.05$ and $||a_D|-1| < 0.05$.
Right: same, with $\cos \alpha_1$ replaced by sgn$(C) \, a_D$.}
\label{fig3}
\end{figure}

We will now discuss the behaviour of $\cos \alpha_1$ (figure~\ref{fig3} left)
and sgn$(C) \, a_D$ (figure~\ref{fig3} right) as a function of $\phi_D = \phi_\tau$ for Type II.
In figure~\ref{fig3} $\sqrt{s}=13$ TeV and all rates are taken at $10$\% (blue/black); 
in red/dark-grey we show the points with $|a_D| < 0.1$ and $||b_D|-1| < 0.1$ and
in green $|b_D| < 0.05$ and $||a_D|-1| < 0.05$. 
The SM-like scenario sgn$(C)$ $(a_D, \, b_D ) = (1,0)$ is easily distinguishable
from the $(0,1)$ scenario. In fact, a measurement of $\phi_\tau$ even if not very precise would easily
exclude one of the scenarios. Obviously, all other scenarios in between these two will need
more precision (and other measurements) to find the values of scalar and pseudoscalar components.
The $\tau^+ \tau^-  h$ angle is related to $\alpha_2$ as
\begin{equation}
 \tan \phi_\tau  = - s_\beta/c_1 \, \tan \alpha_2 \qquad  \Rightarrow \qquad
 \tan \alpha_2 = - c_1/s_\beta \, \tan \phi_\tau
\end{equation}
and therefore a measurement of the angle $\phi_\tau$ does not directly constrain
the angle $\alpha_2$ but rather a relation between the three angles.
A measurement of $\phi_t$ and $\phi_\tau$ would give us two independent relations
to determine the three angles.

\subsection{Constraints from EDM}

The C2HDM, as all models with CP violation, are constrained 
by bounds arising from the measurement of electric dipole 
moments (EDMs) of neutrons, atoms and molecules. 
The parameter space of the C2HDM was analysed in~\cite{Buras:2010zm, 
Cline:2011mm, Jung:2013hka, Shu:2013uua, Inoue:2014nva, Brod:2013cka} and
ref.~\cite{Inoue:2014nva} found that the most stringent bounds are 
obtained using the results from the ACME Collaboration~\cite{Baron:2013eja}, 
except when cancellations among the neutral scalars occur. These cancellations were pointed 
out in~\cite{Jung:2013hka, Shu:2013uua} and arise due to orthogonality of 
the $R$ matrix in the case of almost degenerate scalars~\cite{Fontes:2014xva}. 

The scenarios we discuss have the 
couplings of the up-type sector (top quark) very close to the SM ones.
Indeed, only the couplings in the down-type sector, namely tau lepton
and the b-quark Yukawa couplings, are still allowed by data to have a vanishing 
scalar component. Due to the universality of the lepton Yukawa couplings 
the electron EDM also restricts the tau Yukawa. In a type II model this in turn
also restricts the b-quark Yukawa of the SM-like Higgs. However, this is 
completely irrelevant in the 
Flipped model, where the charged leptons couple as the up-type quarks.
We have shown in~\cite{Fontes:2015mea} that in a preliminary scan over the parameter space we have 
found points which pass all constraints including ACME's.

A dedicated study of the EDM contributions in the Type II C2HDM, where there 
are several sources of CP violation and where the partial cancellations of 
the various scalars is dully taken into account is in progress~\cite{us_edm}. 
In addition, one should keep in mind that, as pointed out 
in ref.~\cite{Brod:2013cka, Dekens:2014jka}, the future bounds from the EDMs can have a 
strong impact on the C2HDM. In the future, the interplay between the EDM bounds and the 
data from the LHC Run 2 will pose relevant new constraints in the complex 
2HDM in general, and in particular for the scenarios presented in this 
work.

\section{Conclusions}
\label{sec:conc}

We have discussed the interesting possibility of having a vanishing scalar component
is some of the Yukawa couplings, namely the couplings of the lightest Higgs to 
down-type quarks and/or to leptons.
These scenarios can occur for Type II, F and LS, and the pseudoscalar component
plays the role of the scalar component in assuring the measured rates at the LHC.
A direct measurement of the angles that gauge the ratio of pseudoscalar to scalar
components is needed to further constrain the model. In particular, the measurement of $\phi_\tau$, 
the angle for the $\tau^+ \tau^- h$ vertex, will allow to either confirm or to rule out 
the scenario of a vanishing scalar,
even with a poor accuracy. We have also noted that for the Type F, only a direct
measurement of $\phi_D$ in a process involving the $bbh$ vertex would probe
the vanishing scalar scenario. Finally a future linear collider~\cite{Ono:2012ah, Asner:2013psa} 
will certainly help to further probe the vanishing scalar scenarios.

\bigskip 

\begin{thebibliography}{99}


\bibitem{ATLASHiggs}
G.~Aad {\it et al.}  [ATLAS Collaboration],
  Phys.\ Lett.\ B {\bf 716}, 1 (2012)
  [arXiv:1207.7214 [hep-ex]].

\bibitem{CMSHiggs}
S.~Chatrchyan {\it et al.}  [CMS Collaboration],
  Phys.\ Lett.\ B {\bf 716}, 30 (2012)
  [arXiv:1207.7235 [hep-ex]].


\bibitem{Lee:1973iz}
 T.D.~Lee,
 Phys.\ Rev.\  D {\bf 8} (1973) 1226.




\bibitem{hhg}
  J.F.~Gunion, H.E.~Haber, G.L.~Kane and S.~Dawson,
  \textit{The Higgs Hunter's Guide}
  \mbox{(Westview Press, Boulder, CO, 2000)}.

\bibitem{ourreview}
G.~C.~Branco, P.~M.~Ferreira, L.~Lavoura, M.~N.~Rebelo, M.~Sher, and J.~P.~Silva,
\emph{Theory and phenomenology of two-Higgs-doublet models},
\emph{Phys.\ Rept.\ }  {\bf 516} (2012) 1
[arXiv:1106.0034 [hep-ph]].


\bibitem{Barroso:2012wz}
  A.~Barroso, P.~M.~Ferreira, R.~Santos and J.~P.~Silva,
  Phys.\ Rev.\ D {\bf 86} (2012) 015022
  [arXiv:1205.4247 [hep-ph]].

\bibitem{Inoue:2014nva}
  S.~Inoue, M.~J.~Ramsey-Musolf and Y.~Zhang,
  Phys.\ Rev.\ D {\bf 89} (2014) 115023
  [arXiv:1403.4257 [hep-ph]].

\bibitem{Cheung:2014oaa}
  K.~Cheung, J.~S.~Lee, E.~Senaha and P.~Y.~Tseng,
  JHEP {\bf 1406} (2014) 149
  [arXiv:1403.4775 [hep-ph]].

\bibitem{Fontes:2014xva}
  D.~Fontes, J.~C.~Rom\~ao and J.~P.~Silva,
  JHEP {\bf 1412}, 043 (2014)
  [arXiv:1408.2534 [hep-ph]].



\bibitem{Brod:2013cka}
  J.~Brod, U.~Haisch and J.~Zupan,
  JHEP {\bf 1311} (2013) 180
  [arXiv:1310.1385 [hep-ph], arXiv:1310.1385].


\bibitem{ATLASpred}
ATLAS Collaboration, Physics at a High-Luminosity LHC with ATLAS, (2013), arXiv:1307.7292
[hep-ex], SNOW13-00078; ATLAS Collaboration, Projections for measurements of Higgs boson cross sections, branching ratios
and coupling parameters with the ATLAS detector at a HL-LHC, ATLAS-PHYS-PUB-2013-014 (2013).

\bibitem{CMSpred}
CMS Collaboration, Projected Performance of an Upgraded CMS Detector at the LHC and HL-LHC:
Contribution to the Snowmass Process, (2013), arXiv:1307.7135 [hep-ex], SNOW13-00086.


\bibitem{Dawson:2013bba}
  S.~Dawson, A.~Gritsan, H.~Logan, J.~Qian, C.~Tully, R.~Van Kooten, A.~Ajaib and A.~Anastassov {\it et al.},
  arXiv:1310.8361 [hep-ex].



\bibitem{Fontes:2015mea}
  D.~Fontes, J.~C.~Rom\~ao, R.~Santos and J.~P.~Silva,
  JHEP, to be published, arXiv:1502.01720 [hep-ph].

\bibitem{Ginzburg:2002wt}
I.~F.~Ginzburg, M.~Krawczyk and P.~Osland,
\emph{Two Higgs doublet models with CP violation},
hep-ph/0211371.

\bibitem{Khater:2003wq}
W.~Khater and P.~Osland,
\emph{CP violation in top quark production at the LHC and two Higgs doublet models},
\emph{Nucl.\ Phys.\ B} \textbf{661} (2003) 209 [hep-ph/0302004].

\bibitem{ElKaffas:2007rq}
A.~W.~El Kaffas, P.~Osland and O.~M.~Ogreid,
\emph{CP violation, stability and unitarity of the two Higgs doublet model},
\emph{Nonlin.\ Phenom.\ Complex Syst.}
\textbf{10}(2007) 347 [hep-ph/0702097].

\bibitem{ElKaffas:2006nt}
A.~W.~El Kaffas, W.~Khater, O.~M.~Ogreid, and P.~Osland,
\emph{Consistency of the two Higgs doublet model and CP violation in
top production at the LHC},
\emph{Nucl.\ Phys.\ B} \textbf{775} (2007) 45 [hep-ph/0605142].

\bibitem{Grzadkowski:2009iz}
B.~Grzadkowski and P.~Osland,
\emph{Tempered Two-Higgs-Doublet Model},
\emph{Phys.\ Rev.\ D} {\bf 82} (2010) 125026
[arXiv:0910.4068 [hep-ph]].

\bibitem{Arhrib:2010ju}
A.~Arhrib, E.~Christova, H.~Eberl and E.~Ginina,
\emph{CP violation in charged Higgs production and decays in
the Complex Two Higgs Doublet Model},
\emph{JHEP} {\bf 1104} (2011) 089
[arXiv:1011.6560 [hep-ph]].


\bibitem{LS}
L.\ Lavoura, J.\ P.\ Silva,
\emph{Fundamental CP violating quantities in a
$SU(2) x U(1)$ model with many Higgs doublets},
\emph{Phys.\ Rev.\ D} \textbf{50} (1994) 4619 [hep-ph/9404276].

\bibitem{BS}
F.\ J.\ Botella and  J.\ P.\ Silva,
\emph{Jarlskog - like invariants for theories with scalars and fermions},
\emph{Phys.\ Rev.\ D} \textbf{51} (1995) 3870 [hep-ph/9411288].
%


 \bibitem{GWP}
S.L.~Glashow and S.~Weinberg,
Phys.\ Rev.\ D {\bf 15}, 1958 (1977);
E.A.~Paschos,
Phys.\ Rev.\ D {\bf 15}, 1966 (1977).





\bibitem{Aad:2014aba}
  G.~Aad {\it et al.}  [ATLAS Collaboration],
  Phys.\ Rev.\ D {\bf 90} (2014) 5,  052004
  [arXiv:1406.3827 [hep-ex]].

\bibitem{Khachatryan:2014jba}
  V.~Khachatryan {\it et al.}  [CMS Collaboration],
  arXiv:1412.8662 [hep-ex].

\bibitem{BB}
  T.~Hermann, M.~Misiak and M.~Steinhauser,
  JHEP {\bf 1211} (2012) 036
  [arXiv:1208.2788 [hep-ph]];
    F.~Mahmoudi and O.~Stal,
  Phys.\ Rev.\ D {\bf 81}, 035016 (2010)
  [arXiv:0907.1791 [hep-ph]].

\bibitem{Deschamps:2009rh}
  O.~Deschamps, S.~Descotes-Genon, S.~Monteil, V.~Niess, S.~T'Jampens and V.~Tisserand,
  Phys.\ Rev.\ D {\bf 82}, 073012 (2010)
  [arXiv:0907.5135 [hep-ph]].

\bibitem{gfitter1}
M.~Baak, M.~Goebel, J.~Haller, A.~Hoecker, D.~Ludwig,
K.~Moenig, M.~Schott and J.~Stelzer,
\emph{Updated Status of the Global Electroweak Fit and Constraints
on New Physics},
\emph{Eur.\ Phys.\ J.\ C} \textbf{72} (2012) 2003
[arXiv:1107.0975 [hep-ph]]

\bibitem{Ztobb}
  A.~Denner, R.J.~Guth, W.~Hollik and J.H.~Kuhn,
  Z.\ Phys.\ C {\bf 51}, 695 (1991);
%
%
%
  H.E.~Haber and H.E.~Logan,
  Phys.\ Rev.\ D {\bf 62}, 015011 (2000)
  [hep-ph/9909335];
  A.~Freitas and Y.-C.~Huang,
  JHEP {\bf 1208}, 050 (2012)
  [arXiv:1205.0299 [hep-ph]].

\bibitem{Abbiendi:2013hk}
  G.~Abbiendi {\it et al.}  [ALEPH and DELPHI and L3 and OPAL and LEP Collaborations],
  Eur.\ Phys.\ J.\ C {\bf 73} (2013) 2463
  [arXiv:1301.6065 [hep-ex]].

\bibitem{ATLASICHEP}
  ATLAS collaboration, ATLAS-CONF-2013-090;
  G.~Aad {\it et al.}  [ATLAS Collaboration],
  JHEP {\bf 1206} (2012) 039
  [arXiv:1204.2760 [hep-ex]];
   G.~Aad {\it et al.}  [ATLAS Collaboration],
  arXiv:1412.6663 [hep-ex].

\bibitem{CMSICHEP}
  S.~Chatrchyan {\it et al.}  [CMS Collaboration],
  JHEP {\bf 1207} (2012) 143
  [arXiv:1205.5736 [hep-ex]]; CMS Note, CMS-PAS-HIG-14-020.

\bibitem{lepewwg}
The ALEPH, CDF,  D0, DELPHI, L3, OPAL, SLD Collaborations, the LEP Electroweak Working Group, the Tevatron Electroweak Working Group, and the SLD electroweak and heavy flavour Groups,
  arXiv:1012.2367 [hep-ex].

\bibitem{ATLAS:2012ac}
  G.~Aad {\it et al.}  [ATLAS Collaboration],
  Phys.\ Lett.\ B {\bf 710} (2012) 383
  [arXiv:1202.1415 [hep-ex]].
  
\bibitem{Chatrchyan:2013yoa}
  S.~Chatrchyan {\it et al.}  [CMS Collaboration],
  Eur.\ Phys.\ J.\ C {\bf 73} (2013) 2469
  [arXiv:1304.0213 [hep-ex]].


\bibitem{Aad:2014vgg}
  G.~Aad {\it et al.}  [ATLAS Collaboration],
  JHEP {\bf 1411} (2014) 056
  [arXiv:1409.6064 [hep-ex]].
  
  
\bibitem{Khachatryan:2014jya}
  V.~Khachatryan {\it et al.}  [CMS Collaboration],
  Phys.\ Rev.\ D {\bf 90} (2014) 112013
  [arXiv:1410.2751 [hep-ex]].



\bibitem{Deshpande:1977rw}
N.~G.~Deshpande and E.~Ma,
\emph{Pattern of Symmetry Breaking with Two Higgs Doublets},
\emph{Phys.\ Rev.\ D} \textbf{18} (1978) 2574 .

\bibitem{Kanemura:1993hm}
S.~Kanemura, T.~Kubota and E.~Takasugi,
\emph{Lee-Quigg-Thacker bounds for Higgs boson masses in a two doublet model},
\emph{Phys.\ Lett.\ B} \textbf{313} (1993) 155
[hep-ph/9303263].

\bibitem{Akeroyd:2000wc}
A.~G.~Akeroyd, A.~Arhrib and E.~-M.~Naimi,
\emph{Note on tree level unitarity in the general two Higgs doublet model},
\emph{Phys.\ Lett.\ B} \textbf{490}(2000) 119
[hep-ph/0006035].

\bibitem{Ginzburg:2003fe}
I.~F.~Ginzburg and I.~P.~Ivanov,
\emph{Tree level unitarity constraints in the 2HDM with CP violation},
hep-ph/0312374.

\bibitem{Peskin:1991sw}
  M.E.~Peskin and T.~Takeuchi,
  Phys.\ Rev.\ D {\bf 46}, 381 (1992).

\bibitem{Grimus:2008nb}
W.~Grimus, L.~Lavoura, O.~M.~Ogreid and P.~Osland,
\emph{The Oblique parameters in multi-Higgs-doublet models}
\emph{Nucl.\ Phys.\ B} \textbf{801} (2008) 81
[arXiv:0802.4353 [hep-ph]].

\bibitem{Baak:2012kk}
M.~Baak, M.~Goebel, J.~Haller, A.~Hoecker, D.~Kennedy, R.~Kogler, K.~Moenig and M.~Schott et al.,
\emph{The Electroweak Fit of the Standard Model after the
Discovery of a New Boson at the LHC},
\emph{Eur.\ Phys.\ J.\ C} \textbf{72} (2012) 2205
[arXiv:1209.2716 [hep-ph]].




\bibitem{Spira:1995mt}
M.~Spira,
\emph{HIGLU: A program for the calculation of the total
Higgs production cross-section at hadron colliders via gluon
fusion including QCD corrections},
hep-ph/9510347.

\bibitem{Harlander:2012pb}
R.~V.~Harlander, S.~Liebler and H.~Mantler,
\emph{SusHi: A program for the calculation of Higgs production in
gluon fusion and bottom-quark annihilation in the Standard Model
and the MSSM},
\emph{Comput. Phys. Commun.} \textbf{184} (2013) 1605
[arXiv:1212.3249 [hep-ph]].

\bibitem{LHCCrossSections}
https://twiki.cern.ch/twiki/bin/view/LHCPhysics/CrossSectionsFigures .
%

\bibitem{Fontes:2014tga}
  D.~Fontes, J.~C.~Rom\~ao and J.~P.~Silva,
  Phys.\ Rev.\ D {\bf 90}, no. 1, 015021 (2014)
  [arXiv:1406.6080 [hep-ph]].

\bibitem{Ferreira:2014dya}
  P.~M.~Ferreira, R.~Guedes, M.~O.~P.~Sampaio and R.~Santos,
  JHEP {\bf 1412} (2014) 067
  [arXiv:1409.6723 [hep-ph]].


\bibitem{Gunion:1996xu}
  J.~F.~Gunion and X.~G.~He,
  Phys.\ Rev.\ Lett.\  {\bf 76} (1996) 4468
  [hep-ph/9602226].


\bibitem{Ellis:2013yxa}
  J.~Ellis, D.~S.~Hwang, K.~Sakurai and M.~Takeuchi,
  JHEP {\bf 1404} (2014) 004
  [arXiv:1312.5736 [hep-ph]].

\bibitem{He:2014xla}
  X.~G.~He, G.~N.~Li and Y.~J.~Zheng,
  arXiv:1501.00012 [hep-ph].

\bibitem{Boudjema:2015nda}
  F.~Boudjema, R.~M.~Godbole, D.~Guadagnoli and K.~A.~Mohan,
  arXiv:1501.03157 [hep-ph].


\bibitem{DelDuca:2001fn}
  V.~Del Duca, W.~Kilgore, C.~Oleari, C.~Schmidt and D.~Zeppenfeld,
  Nucl.\ Phys.\ B {\bf 616} (2001) 367
  [hep-ph/0108030].

\bibitem{Field:2002gt}
  B.~Field,
  Phys.\ Rev.\ D {\bf 66} (2002) 114007
  [hep-ph/0208262].

\bibitem{Dolan:2014upa}
  M.~J.~Dolan, P.~Harris, M.~Jankowiak and M.~Spannowsky,
  Phys.\ Rev.\ D {\bf 90} (2014) 7,  073008
  [arXiv:1406.3322 [hep-ph]].



\bibitem{Berge:2008wi}
  S.~Berge, W.~Bernreuther and J.~Ziethe,
  Phys.\ Rev.\ Lett.\  {\bf 100} (2008) 171605
  [arXiv:0801.2297 [hep-ph]].


\bibitem{Harnik:2013aja}
  R.~Harnik, A.~Martin, T.~Okui, R.~Primulando and F.~Yu,
  Phys.\ Rev.\ D {\bf 88} (2013) 7,  076009
  [arXiv:1308.1094 [hep-ph]].

\bibitem{Askew:2015mda}
  A.~Askew, P.~Jaiswal, T.~Okui, H.~B.~Prosper and N.~Sato,
  arXiv:1501.03156 [hep-ph].
  

\bibitem{Berge:2014sra}
  S.~Berge, W.~Bernreuther and S.~Kirchner,
  Eur.\ Phys.\ J.\ C {\bf 74} (2014) 11,  3164
  [arXiv:1408.0798 [hep-ph]].


\bibitem{Buras:2010zm}
  A.~J.~Buras, G.~Isidori and P.~Paradisi,
  Phys.\ Lett.\ B {\bf 694}, 402 (2011)
  [arXiv:1007.5291 [hep-ph]].

\bibitem{Cline:2011mm}
  J.~M.~Cline, K.~Kainulainen and M.~Trott,
  JHEP {\bf 1111}, 089 (2011)
  [arXiv:1107.3559 [hep-ph]].

\bibitem{Jung:2013hka}
  M.~Jung and A.~Pich,
  JHEP {\bf 1404}, 076 (2014)
  [arXiv:1308.6283 [hep-ph]].

\bibitem{Shu:2013uua}
  J.~Shu and Y.~Zhang,
  Phys.\ Rev.\ Lett.\  {\bf 111}, no. 9, 091801 (2013)
  [arXiv:1304.0773 [hep-ph]].


\bibitem{Baron:2013eja}
  J.~Baron {\it et al.}  [ACME Collaboration],
  Science {\bf 343}, 269 (2014)
  [arXiv:1310.7534 [physics.atom-ph]].


\bibitem{us_edm}  D.~Fontes, J.~C.~Rom\~ao, 
R.~Santos and J.~P.~Silva,  \textit{to appear}. 

\bibitem{Dekens:2014jka}
  W.~Dekens, J.~de Vries, J.~Bsaisou, W.~Bernreuther, C.~Hanhart, U.~G.~Mei{\ss}ner, A.~Nogga and A.~Wirzba,
  JHEP {\bf 1407} (2014) 069
  [arXiv:1404.6082 [hep-ph]].




\bibitem{Ono:2012ah}
  H.~Ono and A.~Miyamoto,
  Eur.\ Phys.\ J.\ C {\bf 73} (2013) 2343.

\bibitem{Asner:2013psa}
  D.M.~Asner, T.~Barklow, C.~Calancha, K.~Fujii, N.~Graf, H.E.~Haber, A.~Ishikawa, S.~Kanemura {\it et al.},
  arXiv:1310.0763 [hep-ph].


\end{thebibliography}
\begin{acknowledgments}
%
%
RS is grateful to the workshop organisation for financial support and for providing the 
opportunity for very stimulating discussions.
RS is supported in part by the Portuguese
\textit{Funda\c{c}\~{a}o para a Ci\^{e}ncia e Tecnologia} (FCT)
under contract PTDC/FIS/117951/2010.
DF, JCR and JPS are also supported by the Portuguese Agency FCT
under contracts
CERN/FP/123580/2011,
EXPL/FIS-NUC/0460/\-2013 and PEst-OE/FIS/UI0777/2013.
\end{acknowledgments}

\end{document}